\documentclass[preprint]{elsarticle}
\usepackage[top=0.25in, bottom=0.25in, left=0.5in, right=0.5in]{geometry}

\usepackage{amssymb}
\usepackage{amsmath}

\journal{Nuclear Physics B}

\begin{document}

\begin{frontmatter}



\title{Development of a Custom kV-amplitude, pressure-tolerant Radio-Frequency  transmitter} 


\author{Christian Hornhuber, Mohammad Ful Hossain Seikh, Mark Stockham, Scott Voigt, Rob Young, Alisa Nozdrina, Sanyukta Agarwal, Shoukat Ali, Kenny Couberly and Dave Besson} 

\affiliation{organization={University of Kansas},
            addressline={1082 Malott Hall}, 
            city={Lawrence},
            postcode={66045}, 
            state={Kansas},
            country={USA}}

\begin{abstract}
Current experiments seeking first-ever observation of Ultra-High Energy Neutrinos (UHEN) typically utilize radio frequency (RF) receiver antennas deployed in cold, radio-transparent polar ice, to measure the coherent RF signals resulting from neutrino interactions with ice molecules. Accurate calibration of the receiver response, sampling  the full range of possible neutrino geometries,
is necessary to estimate the energy and incoming direction of the incident neutrino.
Herein, we detail the design and performance of a custom radio-frequency calibration transmitter, consisting of a battery-powered, kiloVolt-scale signal generator (`IDL' pulser) driving an antenna (South Pole UNiversity of Kansas antenna, or `SPUNK') capable of operating at pressures of 200 atmospheres. Performance was validated by lowering the transmitter into a borehole at the South Pole to a depth of 1740 m, yielding high Signal-to-Noise ratio signals at a distance of 5 km from the source. 
\end{abstract}




\end{frontmatter}




\section{Introduction}
\label{sec1}

Ultra-high energy neutrinos (UHEN) can be detected by measuring radio signals induced by hadronic or electromagnetic showers developing in dense media. The electromagnetic particle cascade initiated by UHEN consists of electrons, positrons and photons. As pre-existing in-ice atomic electrons deplete the positron population in the shower via annihilation, and are also Compton scattered by shower photons into the shower itself, the cascade acquires a net negative charge. This time-varying net-excess charge creates Cherenkov radiation which is coherent for wavelengths greater than the transverse scale of the shower ($\sim$10 cm), the so-called `Askaryan effect' \citep{Askaryan:1961,Askaryan1965CoherentRE}.

Following initial simulations in the early 90's by Zas, Halzen and Stanev and also Frichter, McKay and Ralston \citep{zas1992electromagnetic,Frichter:1995cn}, 25-cm length, `fat-dipole' antennas were deployed into dry holes at 10-200 m depths by the Radio Ice Cherenkov Experiment (RICE) \citep{Kravchenko_2012} at South Pole. This strategy was subsequently followed by the Askaryan Radio Array (ARA) \citep{Allison:2011wk}, the Radio Neutrino Observatory in Greenland (RNO-G)  \citep{RNO-G:2020rmc}, and is planned for the future radio component of the proposed IceCube-Gen2 experiment \citep{Aartsen_2021}. In all these experiments, receiver antennas are deployed within, or just below the firn, with signals conveyed to surface data acquisition either by coaxial cable or optical fiber.

The three decade time span over which this technique has developed notwithstanding, there has been no unambiguous UHEN signal thus far reported. Among the challenges that such experiments must address is identifying a single upcoming, radio-frequency neutrino signal event in a sample of $\sim 10^8$ total event triggers, with backgrounds dominated by down-coming anthropogenic noise generated at the surface, or thermal noise excursions which periodically exceed the trigger threshold. Currently active experiments exploiting this detection strategy at either the South Pole or Greenland employ compact `stations', each consisting of $\sim$16--20 antennas, sensitive over the frequency range 100--500 MHz, deployed over an ice volume of order $8000~m^3$, and designed to detect neutrinos interacting in a volume a factor $10^6\times$ larger.

\begin{figure}
    \centering
    \includegraphics[width=0.6\linewidth]{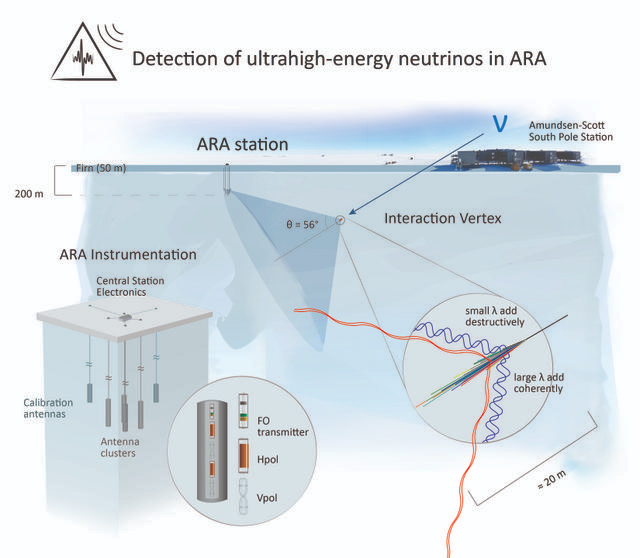}
    \caption{Illustration of neutrino signal generation and detection scheme used by the Askaryan Radio Array experiment at the South Pole. Neutrino ($\nu$) incident from upper left penetrates $\sim$1 km into South Polar ice, and subsequently interacts with ice molecule, resulting in `shower' (zoomed, right); particles in shower produce both incoherent short-wave and coherent longer-wavelength (radio-frequency) radiation, then  observed in 16-antenna ARA array. The goal of the calibration transmitter described here in is to simulate a neutrino interacting in-ice, producing (as in diagram) horizontally propagating RF radiation. Figure reprinted from https://pos.sissa.it/358/933/pdf.}
    \label{fig:pictogram}
\end{figure}

In-ice UHEN experiments, broadly, have two primary science goals. The first is ‘neutrino detection’, for which the reconstruction precision (and therefore the refractive index [$n(z)$] precision, from which it is derived) requirements are significantly less stringent than that required for ’neutrino astronomy’, for which the directions of incident signals must be projected back into the sky. Neutrino detection primarily relies on discriminating neutrino signals originating below the array from anthropogenic backgrounds originating from above the surface. Figure \ref{fig:pictogram} illustrates the detection strategy employed by the Askaryan Radio Array (ARA) at the South Pole, depicting the in-ice collision of a neutrino with an ice molecule, followed by production and subsequent detection of radio-frequency signal.


Since radio-frequency fat dipole receiver antennas are typically deployed in vertically drilled ice boreholes (at 100-200 m depths), signal response is largest in the horizontal plane, corresponding to the direction of maximum elevation gain. Surface transmitter antennas are straightforward to set up and deploy to verify receiver response {\it in situ}, but are therefore geometrically mismatched to dipole receiver beam patterns. Horizontally-propagating calibration transmitter signals, broadcast from sub-surface radio transmitters, can better replicate the typical geometry of actual neutrino signals than calibration transmitters on the surface. Since in-ice neutrino interactions should be detectable to depths of 2--3 km in cold polar ice (either Greenland or Antarctica), calibration transmitters should, ideally, probe similar depths, requiring transmitter designs capable of withstanding the 200 atmosphere pressures to which they would be subjected in an ice borehole filled with drilling fluid (typically an inert organic fluid with a temperature-dependent specific gravity profile well-matched to that of the surrounding ice) at 1-2 km depths. 

Consisting of a signal generator capable of producing fast (ns-scale), high-amplitude (kV-scale) radio-frequency calibration signals in a high-pressure environment, the 'South Pole UNiversity of Kansas Pressure Vessel Antenna' (SPUNK PVA, predecessor to the `Greenland University of Kansas Pressure Vessel Antenna' [GUNK PVA]) was designed to meet these requirements. In this article, we detail the development of the PVA components over the last several years as well as its operational performance in the field.

\section{PVA signal generation with a robust low-jitter and sub-nanosecond high voltage pulser}
\label{subsec1}

The primary active component of the PVA is the signal generator, which must apply kV-amplitude, ns-timescale impulsive voltages at the feedpoint of the transmitter antenna.
Ultra-high speed and high voltage pulsers have found widespread application in medicine, biology and also physics. Our particular application requires signals similar to the large-bandwidth (hundreds of MHz), fast rise-time pulses characteristic of RF generated by cosmic ray interactions with terrestrial matter; such highly resolved pulses can also facilitate measurements of the 
detailed structure and properties of ice in Antarctica and Greenland. Other potential usages include high voltage pulse emissions in low pressure environments, for example low -- orbit satellites or stratospheric balloons also employed in astrophysics experiments.

Such applications necessitate a pulser design which operates stably at low temperatures (-25 to -40°C) and is resistant to spark discharge at low pressure (as would otherwise be expected by P\"aschen's Law for operation at high altitude). In
addition, low-jitter (of order, but not more than ns-scale) signals tied to a high-precision GPS cadence can allow stacking of multiple signals to enhance signal-to-noise ratios. To match the form factor of the ice borehole into which they are deployed, a compact cylindrical design is required, which also
accommodates efficient coupling to an antenna feed point as well as a proximal 
battery pack for power. Typical specifications include a rise time below 1 ns, pulse width below 10 ns and a peak voltage between 800$\to$1500V, driving an effective load impedance of 50 Ohm.

\subsection{Design}
\begin{figure}
    \centering
    \includegraphics[width=\linewidth]{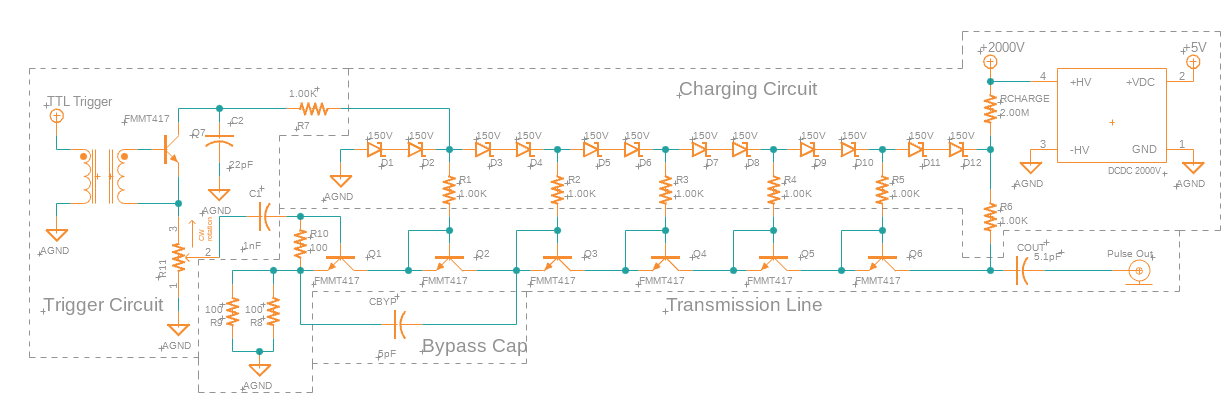}
    \caption{Avalanche Pulser Circuit}
    \label{fig:PulserCircuit}
\end{figure}

Different applications\cite{Avalanche2007,Avalanche2015} discuss the use of avalanche transistors, for which the collector-emitter voltage exceeds the breakdown voltage.
 In the avalanche breakdown mode, the transistor can switch high
currents on sub-nanosecond timescales. The simple solid-state design relies on 
 inexpensive and readily-available components, and is resistant to sparking. 
As shown in Figure \ref{fig:PulserCircuit}, we use the FMMT417 transistor, which is specially designed for avalanche applications. This transistor has a breakdown voltage of 320 V
and can accommodate up to 60 amperes within 20 nanoseconds. Another advantage
of this transistor is the small SOT23-5 package and the small 2.5 nH inductance.
Before building the circuit, we tested all the
transistors and selected components based on uniformity of breakdown voltage and breakdown
fall time. Having similar transistors prevents inconsistent timings of
each breakdown and therefore ensures a constant pulse output.

\subsubsection{Avalanche Transmission Line}
The transmission line consists of 6 avalanche transistors in series, a
charging capacitor, and termination resistors. Each base and emitter are
connected to the collector of the previous transistor, with the exception of Q1 (Fig. \ref{fig:PulserCircuit}),
for which the base is connected to the trigger circuit. The trigger of the
base of Q1 opens its collector -- emitter line, which causes a sharp
voltage gradient at the collector -- emitter at Q2 exceeding the
breakdown voltage of the transistor, inducing Q2 into
avalanche breakdown and fully opening the transistor. This effect
continues with the following transistors, each experiencing successively
higher voltage gradients and therefore faster avalanche breakdowns.
After the last transistor Q6, the charging capacitor COUT discharges and
sends a positive pulse wave through the transistor transmission line and
a negative pulse to the SMA output. Since the load has an effective
impedance of 50 $\Omega$, this value dictates the impedance required of the transmission line, given by

\[Z = \ \sqrt{\frac{L}{C}}\].

In addition to the transistor inductance (2.5 nH), 
we use the 
PCB layout itself to create a small (1 pF) capacitance at each avalanche
transistor to achieve the desired total 50 Ohm impedance. Since the pulse wave is propagating through the stages, it
needs to be terminated (achieved using two 100 Ohm resistors in parallel) after the first transistor Q1 to prevent
reflections in the transmission line. As detailed below, we compared a wide range of values of
the charging capacitor COUT (5 pF $\to$ 220 pF) for our pulser applications; this value controls the characteristic RC constant and therefore influences the width of the generated pulse. 

\subsubsection{Charging Circuit}

Since the breakdown voltage of the avalanche
transistor is 320 V, we must apply a voltage at each
transistor close to the breakdown voltage, but not beyond that (to prevent
premature transistor breakdown). In our application, we chose a
distribution of 300 V at each transistor; the six series stages therefore require a minimum supply of 1800 V. A 2 kV power supply was selected, based on its availability and operational temperature range. In order to ensure equal charging voltages at each transistor, we use two 150 V Zener diodes in series
and supply the transistors through a 1 K$\Omega$ resistor. We also selected a 2 M$\Omega$
 charging resistor RCHARGE, which results in an RC time constant 
\[T \approx 5*2Meg*COUT\]
after each triggered pulse.
\subsubsection{Trigger Circuit}
Pulser operation requires a TTL trigger input from an
external source. Because of the large signals, 
it is desirable to physically isolate the high voltage from the trigger
side using a transformer. In development versions 
of the pulser, for which the base of the transistor Q1 was triggered directly from
the transformer, we observed inconsistency of the pulses as well as 
high jitter relative to the input trigger time (approaching 20 ns). Bench testing traced this to the lack of generated
base current through Q1 by the transformer, which is essential for a
fully open transistor collector -- emitter; only a fully open Q1
collector -- emitter causes a voltage gradient sufficient for the
voltage breakdown of Q2. 

Some applications discuss the use of a pre--pulser stage, which generates significant base current for Q1\cite{Avalanche2015}. Having a
high spike current for the base trigger guarantees pulser triggering 
with low jitter. In our design, we charge a capacitor C2 with 300 V and
discharge it quickly via the avalanche transistor Q7, triggered by the
transformer. The discharge pulse proceeds through a coupling capacitor C1
connected to the base of the first avalanche transistor Q1. A
potentiometer can set the height of the trigger pulse; in our
application we set this to the maximum possible value. Resistor R10 between
Base and Emitter of Q1 sets the transistor avalanche breakdown and has been recommended elsewhere to ensure a stable operation\cite{Avalanche2007,Avalanche2015,ATT2020}.

\subsubsection{Bypass Capacitor}
Development revs of the pulser also exhibited instability (including 
`dropped' pulses) and pulse-to-pulse inconsistency at temperatures below 0 C; at a temperature of --30 °C, the pulser entirely stopped generating pulses. However,
our applications in polar environments require operation down to temperatures of --50 C.
A related cold-temperature problem was the frequent failure of
avalanche transistor Q2 after a minimum period of operation, as noted in \cite{ATT2020}.
An Auxiliary Triggering Topology (ATT) was implemented to enhance reliability. This included the addition of a bypass capacitor CBYP between the base - emitter of Q1 and Q3, which helped to ensure
functionality of Q2, and also restored consistency of the
generated pulses at each trigger. 
Since Q1 is the only
base-triggered transistor, bypassing a small
fraction of the trigger signal to the base of other transistors by a
capacitor serves as a secondary trigger source in addition to the
collector -- emitter breakdown. Previous work advocated use of bypass capacitors
 to almost all transistors, although our testing indicated bypassing only
 the capacitor to Q3 resulted in a satisfactory improvement in performance, including
 preservation of transistor Q2.
 
    According to \cite{ATT2020}, the degradation of Q2 performance 
results from the damaging effects of voltage gradient switching in comparison to trigger-based switching via the leading edge of the transistor voltage. Since Q1 is open longest during the generation of a pulse, it has the highest risk of damage; trigger-based switching of this transistor reduces this risk. Q2, however, which is switched on only by a voltage gradient, has a higher risk of damage.  A bypass capacitor inserted between the Q1 Emitter to the Q2 Collector therefore helps preserve the transistors by bypassing a
portion of the trigger pulse; we found satisfactory results using a 5 pF capacitor (detailed below). 

\subsubsection{Marx Bank Extension Circuit MBC}
\begin{figure}
    \centering
    \includegraphics[width=0.9\linewidth]{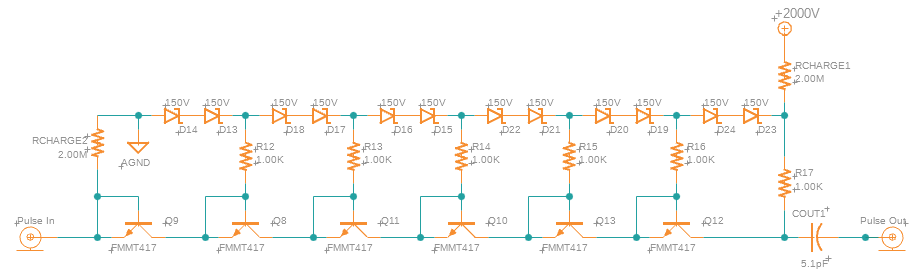}
    \caption{Marx - Bank Extension Circuit}
    \label{fig:MarxBankExtensionCircuit}
\end{figure}

In our application, for which broadcast radio-frequency signals must transit $>$5 km of solid ice before reaching receiver antennas, maximal signal amplitude is desirable. 
We considered two possible ways to increase the output voltage of the pulse. 
The first approach (`brute force') is to simply add more avalanche transistors to the
transmission line; other applications have employed 8 (or more) transistors. Although each additional transistor 
increases the charging voltage by 300 V, this also increases the corresponding
requirement on the high voltage supply. Our use of a 2000 V supply therefore presents
a limit of 6 avalanche transistors for our circuit. The second (more
suitable for our purposes) approach is the usage of a Marx -- Bank circuit (MBC), often used for high voltage pulse generation \cite{Avalanche2015,ATT2020}. An MBC
is a voltage multiplier circuit used in different applications
for high supply voltage generation, consisting of
identical circuits connected with charging capacitors. In
our case, the output of the first circuit is connected to the
first base emitter of the first avalanche transistor of the second
pulser's circuit. The extender pulser circuit in Figure \ref{fig:MarxBankExtensionCircuit} is almost
identical to the first one, modulo the triggering, which is supplied by the 
first stage. In theory, it is possible to add multiple
extender circuits (at the cost of board size) up to the high-voltage DCDC limit.

The Marx -- Bank
circuit works in two steps. The first step is the `steady state' mode, in which
both circuits charge the capacitors. In the second step, the first
circuit is triggered, and creates the initial pulse. This negative pulse
then causes the first transistor of the extender circuit into
avalanche breakdown due to the high negative voltage gradient; the subsequent avalanche then triggers the other
transistors in line. Thus, the second charging capacitor discharges
in series with the first pulser circuit and thereby multiplies not only
the output voltage but also decreases the discharging time by the higher
voltage gradient.

\subsubsection{Physical Layout}

\begin{figure}
    \centering
    \includegraphics[width=0.7\linewidth]{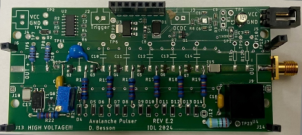}
    \caption{Avalanche Pulser Board}
    \label{fig:AvalanchePulserBoard}
\end{figure}

Our application required a compact pulser layout
(Figure \ref{fig:AvalanchePulserBoard}) given the borehole size constraints, with measurements of 25.4 x 106.7 mm, permitting the pulser to be 
embedded into one chamber of the dipole transmitter antenna (described below) used for calibration. The power supply ranges
from 6 to 24 V, compatible with a compact battery pack, 
and has two TTL inputs for Trigger and DCDC power control,
for energy conserving capabilities.

\begin{figure}
    \centering
    \includegraphics[width=0.7\linewidth]{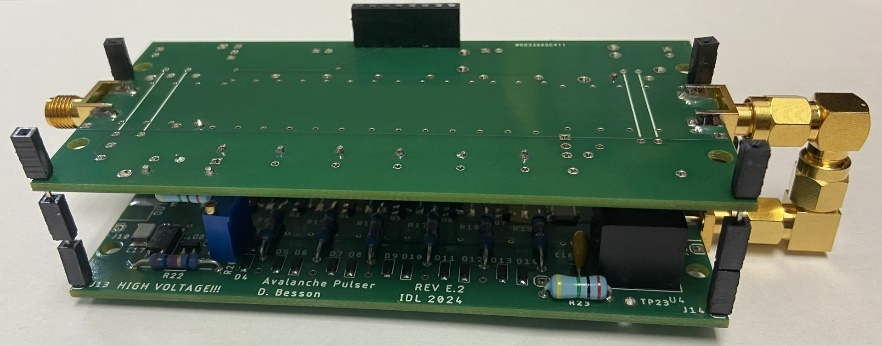}
    \caption{Marx - Bank Pulser Stack}
    \label{fig:MarxBankPulserStack}
\end{figure}

We operate our application mainly in the Marx -- Bank double pulser
extension, with either a series or a parallel stacking of the component boards (Figure \ref{fig:MarxBankPulserStack}). The extension boards are supplied by the main pulser board and the
transmission line is connected via SMA connectors.

\subsection{Performance Measurements}

Measuring high voltage pulses can be a challenge for instrumentation.
For our tests we use three -20dB 50 Ohm attenuators in series between the
pulser and the scope. All have a bandwidth of 6 GHz, well beyond the frequency range
of the pulser itself. For all the measurements we use a scope with a bandwidth of 2.5 GHz and a sample rate of 20 Gs/s.
As discussed earlier, our pulser applications require operation in
temperature environments down to --50 C, requiring testing over a wide
temperature range. We compared
different hardware configurations, as follows:

\subsubsection{Comnparison of single pulser vs Marx -- Bank double pulser with variable charging capacitors}
In general, a generated pulse contains two segments. In the first segment, the avalanche breakdown of the transistors occurs, which leads to an ultra--fast fall of the voltage. As mentioned above, the timing of this rapid fall depends on the negative voltage gradient, with higher voltage gradients giving faster fall times. In theory, the voltage drop of the pulse (pulse amplitude) should be around 50\% of the main charging voltage (900 V in our design). The second part of the pulse is the discharge of the charging capacitor; the discharge time is determined by the termination resistance and by the charging capacitance. Since the double pulser has two charging capacitors in series during pulse generation, the discharge time is half that of the single pulser.

\begin{figure}
    \centering
    \includegraphics[width=0.49\linewidth]{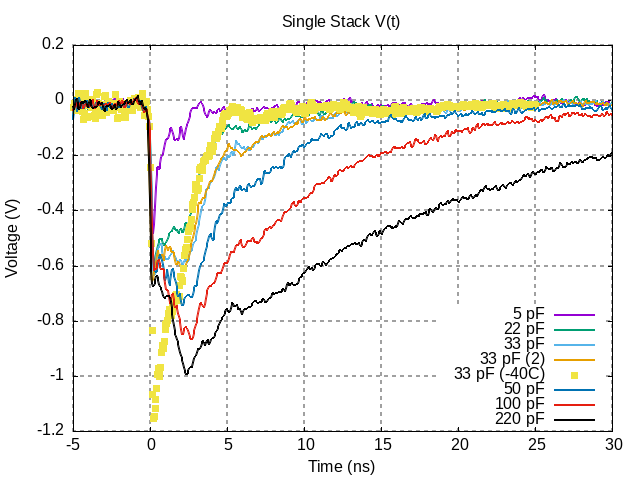}
    \includegraphics[width=0.49\linewidth]{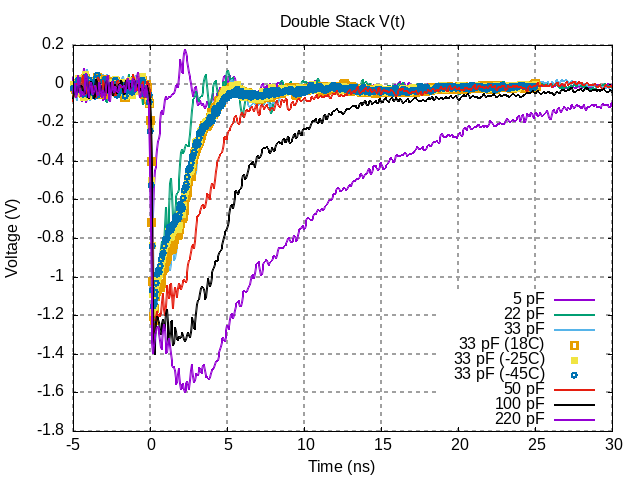}
    \caption{Comparison between single (left) and double pulser signal shapes, as a function of the selected capacitance and, for the case of a 33 pF capacitor, at a variety of tested temperatures. Although all signals have relatively sharp leading edges, we selected the 5 pF variant for production, based on the sharp fall time, in order that the detected signal shape be limited by the bandwidth response of the transmitter/receiver antennas rather than the signal generator. Note the difference in vertical scale between the left and right plots, as well as the relative stability at a variety of tested temperatures in both cases. Measurements were made with a 2.5 GHz oscilloscope, so detection of frequency components much greater than 1 GHz are limited by the scope bandwidth.} 
    \label{fig:SingleDoubleCompareVt}
\end{figure}

In the first test we compared performance of a single pulser vs.
a Marx -- Bank double-stack pulser. Overall, the double-stack pulser not only
achieved a higher output voltage, but also exhibited a faster
fall time due to the higher negative voltage gradient. Figure \ref{fig:SingleDoubleCompareVt} compares signal shapes for a variety of capacitance, and temperature.

The single pulser provides a voltage of around 860 V, a fall time of
around 255 ps and a pulse width of 9.35 ns. Relative to the single-board design, the Marx -- Bank double pulser
achieves a nearly two-fold improvement in peak voltage (1400V), with a fall time of 211 ps and
pulse width of 5 ns. We attribute the fact that the observed pulse voltage is slightly less than a factor of two to i) losses in the transmission line of the pulser, and ii) bandwidth 
limitations of the attenuators, as well as the scope used for recording pulses, given that the 
$\sim$200 ps signal fall time implies signal frequency content beyond the limited bandwidth of the
used scope. 


We infer that the
Marx -- bank double pulser actually has a sharper leading edge than we
can measure with our current instruments. To achieve higher pulse voltages, it is possible to stack several extension pulsers, as described previously, albeit with (slightly) increased trigger jitter at low temperatures.
Numerically, the single pulser jitter (100 ps) increases to a slightly larger, but still very tolerable value of 300 ps at temperatures below -45 C.


\subsubsection{Frequency spectra for different charging capacitors}

As previously discussed, we tested different values of the charging
capacitors (only for the Marx--Bank configuration), ranging from 5 pF to 220 pF, resulting in observed variations in variations of the pulse width.
 Figure \ref{fig:SingleDoubleCompareVt} shows
different generated pulses with different charging capacitor values. As
evident from the Figure, the higher the value of the charging capacitor, the wider the pulse width. Table \ref{tab:SG} compares the results obtained for fall
time, bandwidth, pulse voltage and jitter.

\begin{figure}
    \centering
    \includegraphics[width=0.8\linewidth]{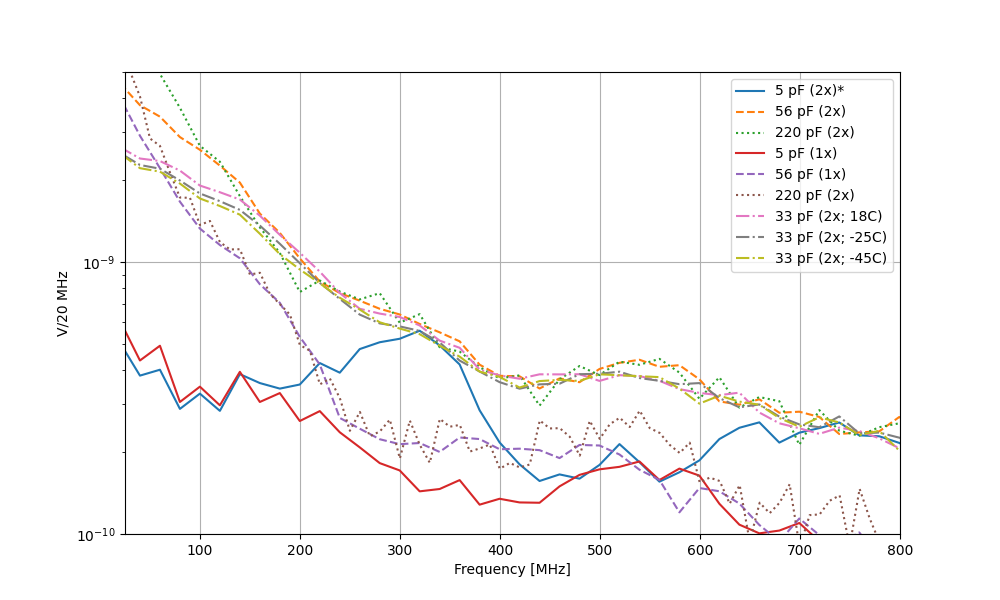}
    \caption{FFT of signal generator signals, as a function of value of discharge capacitor capacitance and also temperature. Value in parenthesis indicates number of boards in pulser. We selected the capacitance value (5 pF, double-stack) that gave the sharpest signal in the time domain, and the most stable operation in the lab.}
    \label{fig:IDLpulsersFFTs.png}
\end{figure}


\begin{table}[h!]
    \centering
    \begin{tabular}{|c|c|c|c|c|}
    \hline
    Capacitor & Pulse Voltage & Fall Time & Pulse Width & Jitter \\
    \hline
    5pF & 850V & 210ps & 0.46ns & 82ps \\
    22pF & 1190V & 203ps & 1.70ns & 118ps \\
    33pf & 1370V & 213ps & 2.10ns & 88ps \\
    56pF & 1350V & 207ps & 3.20ns & 105ps \\
    100pF & 1400V & 211ps & 5.01ns & 86ps \\
    220pF & 1590V & 210ps & 9.41ns & 105ps\\
    \hline
    \end{tabular}
    \caption{Specs of Double Pulser with different charging capacitors}
    \label{tab:SG}
\end{table}

In this comparison, it is notable that the fall time of the pulses and the jitter does
not change significantly. We do, however, observe an increase of the pulse width as we increase the charging capacitor. In theory, the pulse voltage should remain constant, but because of the mentioned bandwidth limitations of attenuators and scope, we are not able to fully measure the pulse shape at lower charging capacitors, leading to the drop of the pulse voltage measured on the bench. This is more obvious when the frequency spectra of these different pulses are compared, displayed in Figure \ref{fig:IDLpulsersFFTs.png}.
At higher frequencies there are no observable differences; we only measure a higher power output at lower frequencies at higher values of the charging capacitor due to the wider pulse width. 

When connected to the transmitter antenna (discussed below), higher values of the charging capacitors lead to unstable behavior of the pulser, resulting from gradual degradation of transistor Q2 with time; we attribute this to the higher current  through the transistor stages with larger capacitors, combined with reflections in the transmission line of the antenna. A charging capacitance of 5 pF has, thus far, resulted in entirely stable operation. 

\section{Antenna Design}
The pulser described above applies a high-voltage signal at the feedpoint of a radio-frequency antenna. The form factor of the pulser itself is therefore dictated by the dimensions and shape of the deployed antenna which, in our case, depends on the dimensions of the ice borehole into which the transmitting antenna is deployed. `Fat dipole' antennas, consisting of opposing conical nose cones welded to longer (typically, a factor of 4--5$\times$) aluminum or copper 80--120 mm diameter tubing provided good frequency response over the desired (200--1000 MHz) range, and were used as the front-end antennas for
 the earliest attempts to detect radio-frequency signals from UHE neutrinos interacting in-ice by the Radio Ice Cherenkov Experiment (RICE).

The high bandwidth offered by the fat dipole model therefore served as the starting point for the development of the antenna used to house the IDL pulser/signal generator described above. Overall, the deployed unit must satisfy the following requirements:
\begin{itemize}
\item Stable pulse generation at temperatures less than -40 C and pressures of up to 200 atm.
\item Output signal strength (combination of signal generator amplitude and antenna gain) adequate to produce a signal-to-noise ratio $>$6 (relative to thermal noise) in a radio receiver at a distance of 5 km.
\item Battery-powered operation for the time period required for an antenna `drop' to 2000 meters and back to the surface (10 hours).
\item Antenna response compatible with Askaryan Radio Array receivers (200-800 MHz).
\end{itemize}
After deciding on the familiar fat dipole design for the transmitter, details of construction (material, dimensions, O-ring selection to ensure pressure seal, etc.) remained; Finite Element Analysis (FEA) codes were used for design optimization. The FEA simulation was conducted assuming a 2500 psi load and requiring a minimum safety factor of 1.7. 

\subsection{Comparison of wall thickness and material variations}
In the early stages of design, the option of having an independent antenna entirely within an RF transparent pressure vessel was considered. It was determined that with the constraints of the OD being smaller than the borehole and the ID being large enough for batteries and electronic components, the wall thickness would have to be too thin for various engineering plastics to ensure structural integrity. Nylon, for example, would have been unsuitable as a pressure vessel material in this application because the FEA indicates that to achieve an acceptable Safety Factor, the tube would need a wall thickness of 0.75 inches, which would reduce the ID of the vessel to 2 inches, inadequate for containing batteries and electronic components. Ultimately, aluminum was selected for the construction material owing to cost, machinability, strength and antenna performance considerations.

Although aluminum tubing with 0.25 inch wall thickness was ultimately used, another potential option was aluminum tubing with 0.125 inch wall thickness, also a common and easy to procure size. The Finite Element Analysis was conducted in the Stress Analysis environment of Autodesk Inventor to determine what wall thickness would provide reliable operation at 2500 psi. The analog for the pressure vessel section of the PVA consisted of a tube of length 19.5 inches and uniform wall thickness, with each end of the tube fixed. The test load was uniformly applied to the outer surface of the tube and set to the maximum value of 2500 psi. Although this would have allowed for more space for the electrical components, the FEA results indicate that the 0.125 inch wall thickness tube would fail with a 2500 psi load (Safety Factor of 0.63). A summary of FEA results are in the table below.

\begin{table}[htpb]
\centering
\begin{tabular}{|cccccc|} \hline
Material & OD (in) & ID (in) & WT (in) & Max Disp (in) & Min Safety Factor \\ \hline
Aluminum & 3.5 & 3.25 & 0.125 & 0.012 & 0.63 \\
Aluminum & 3.5 & 3$^*$ & *0.25 & 0.004 &1.73 \\
Nylon 6/6 &  3.5 & 3 & 0.25 & 0.089 & 0.53 \\
Nylon 6/6 & 3.5 & 2.5 & 0.5 & 0.033 & 0.96 \\
Nylon 6/6 & 3.5 & 2 & 0.75 & 0.016 &  1.72 \\ \hline
\end{tabular}
\caption{Materials considered for construction of SPUNK PVA. Asterisk indicates nominal dimensions for the SPUNK PVA.}
\end{table}

\begin{figure}
    \centering
        \includegraphics[width=0.32\linewidth]{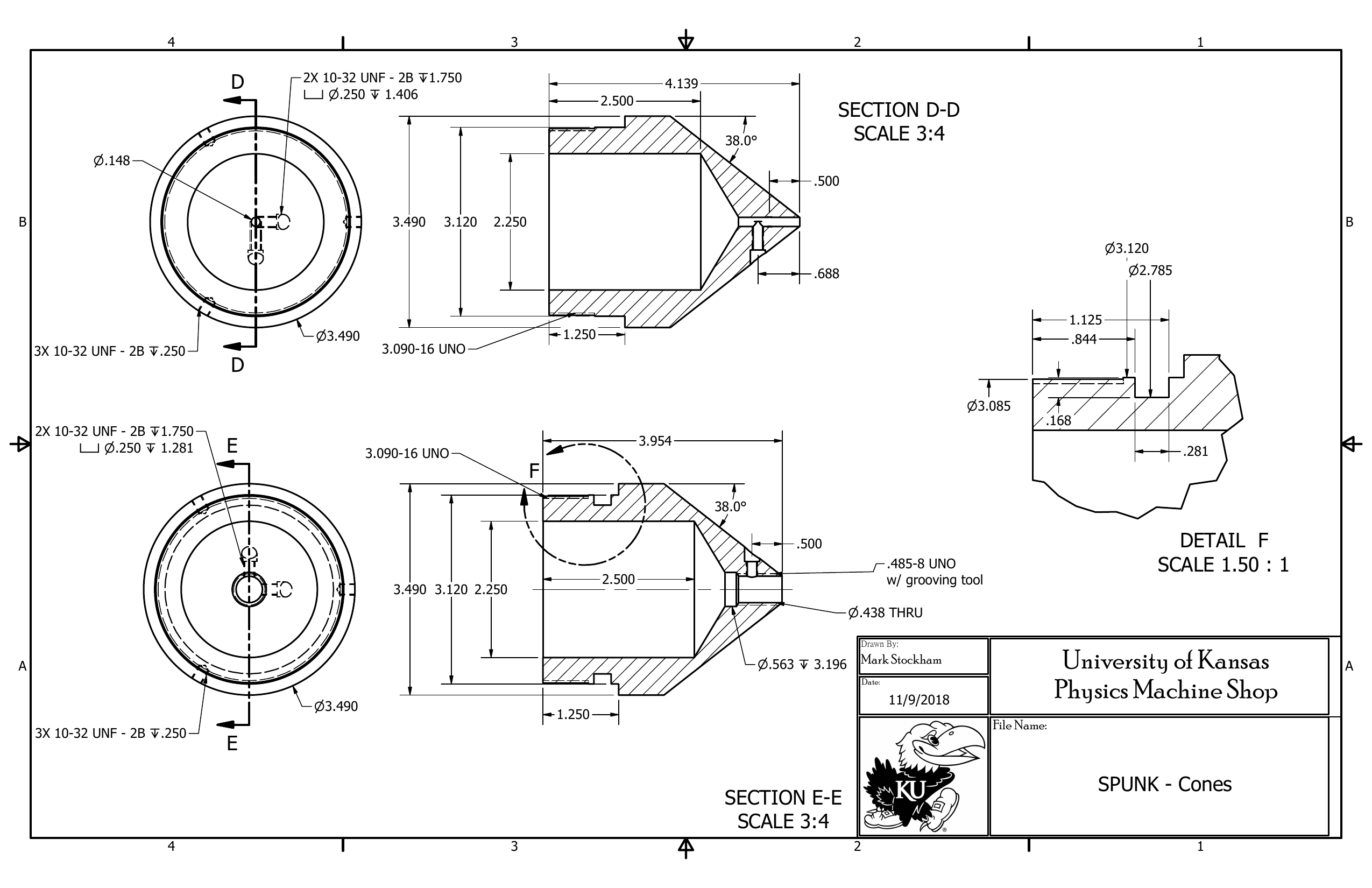}
            \includegraphics[width=0.32\linewidth]{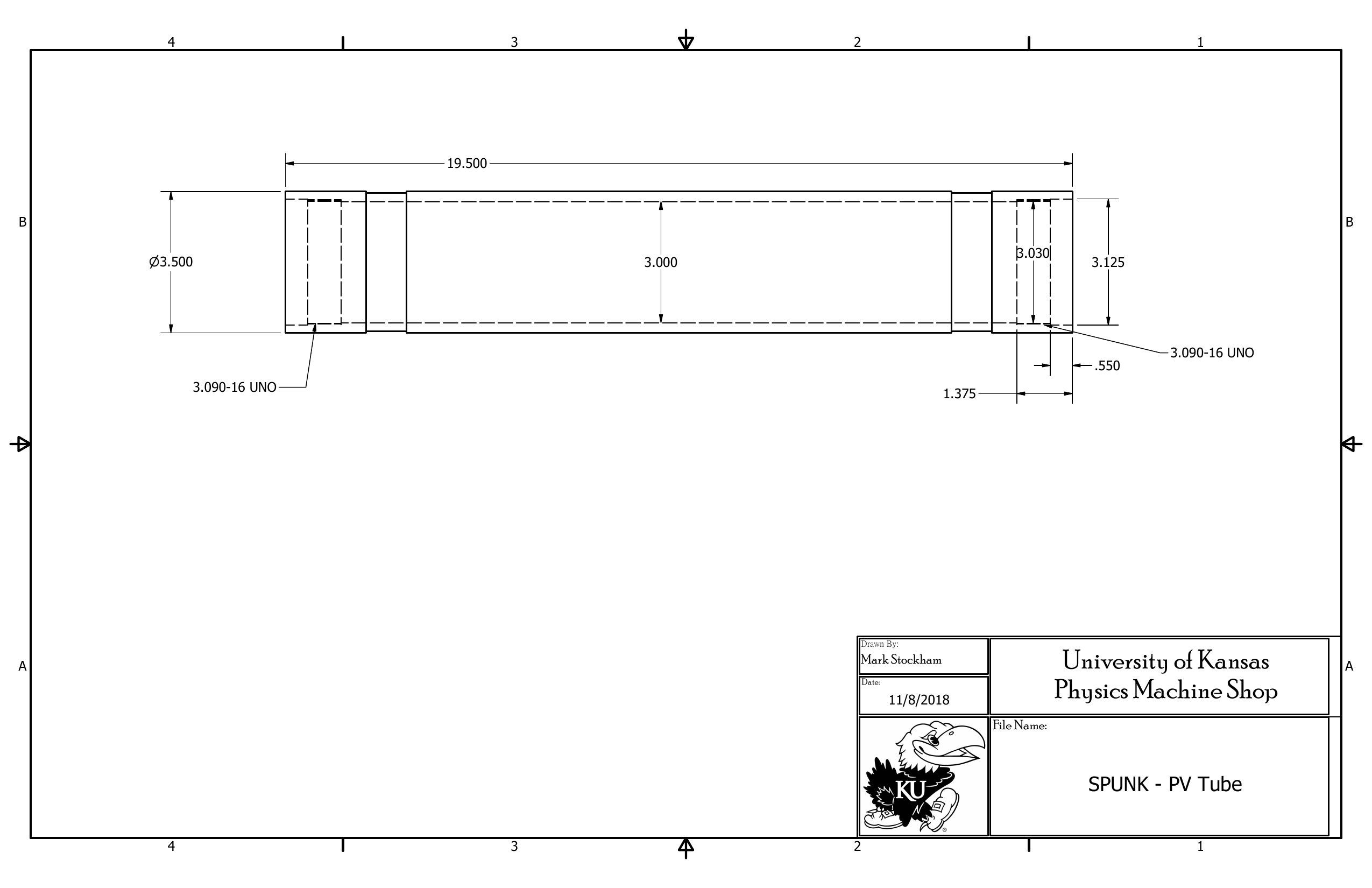}
                \includegraphics[width=0.32\linewidth]{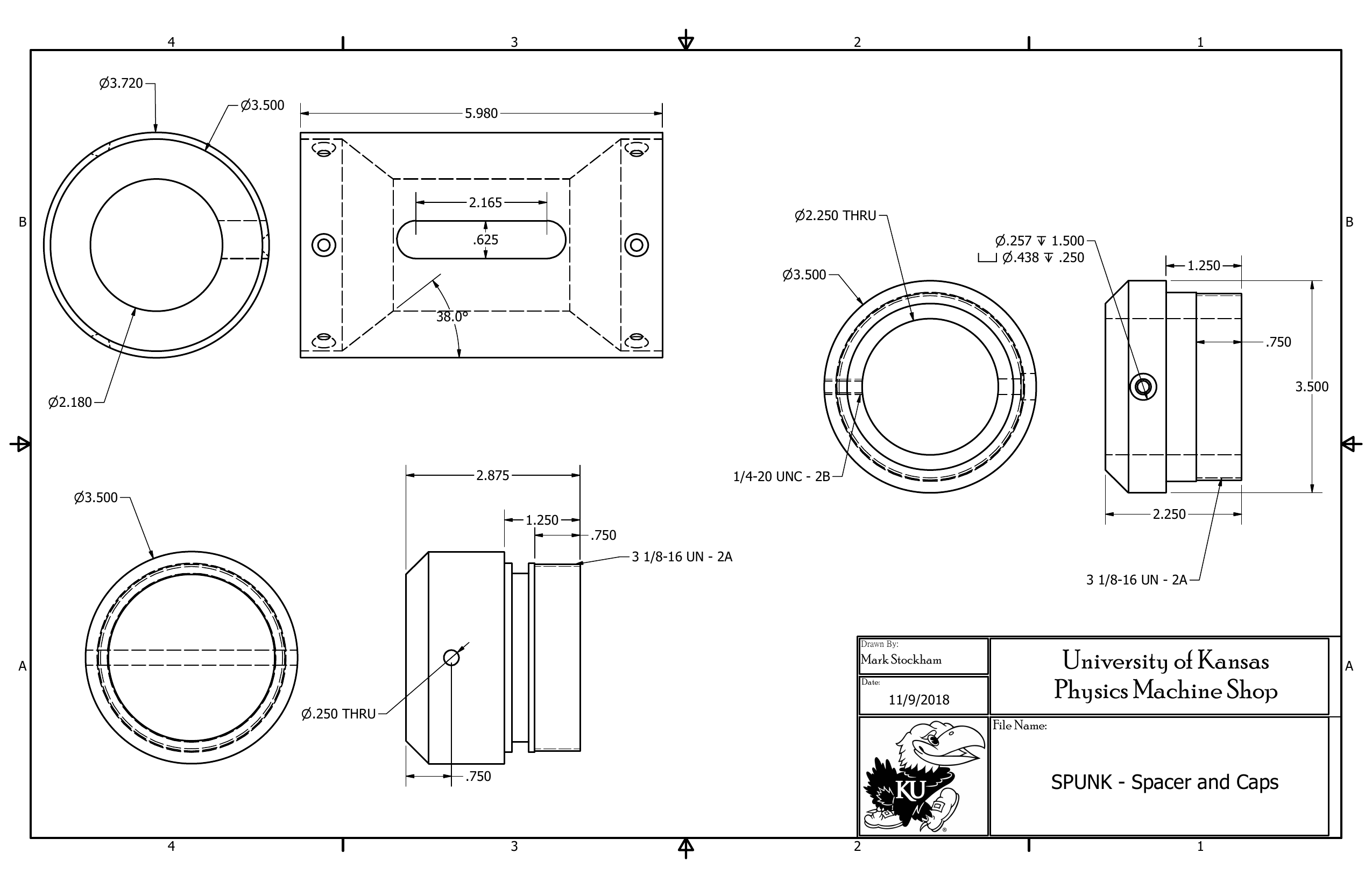}
                \caption{Drawings of nose cones (left), cylindrical barrel (center) and spacer between half-dipoles (right)}
\label{fig:SPUNK_drawings}
\end{figure}

\begin{figure}
    \centering
        \includegraphics[width=0.52\linewidth]{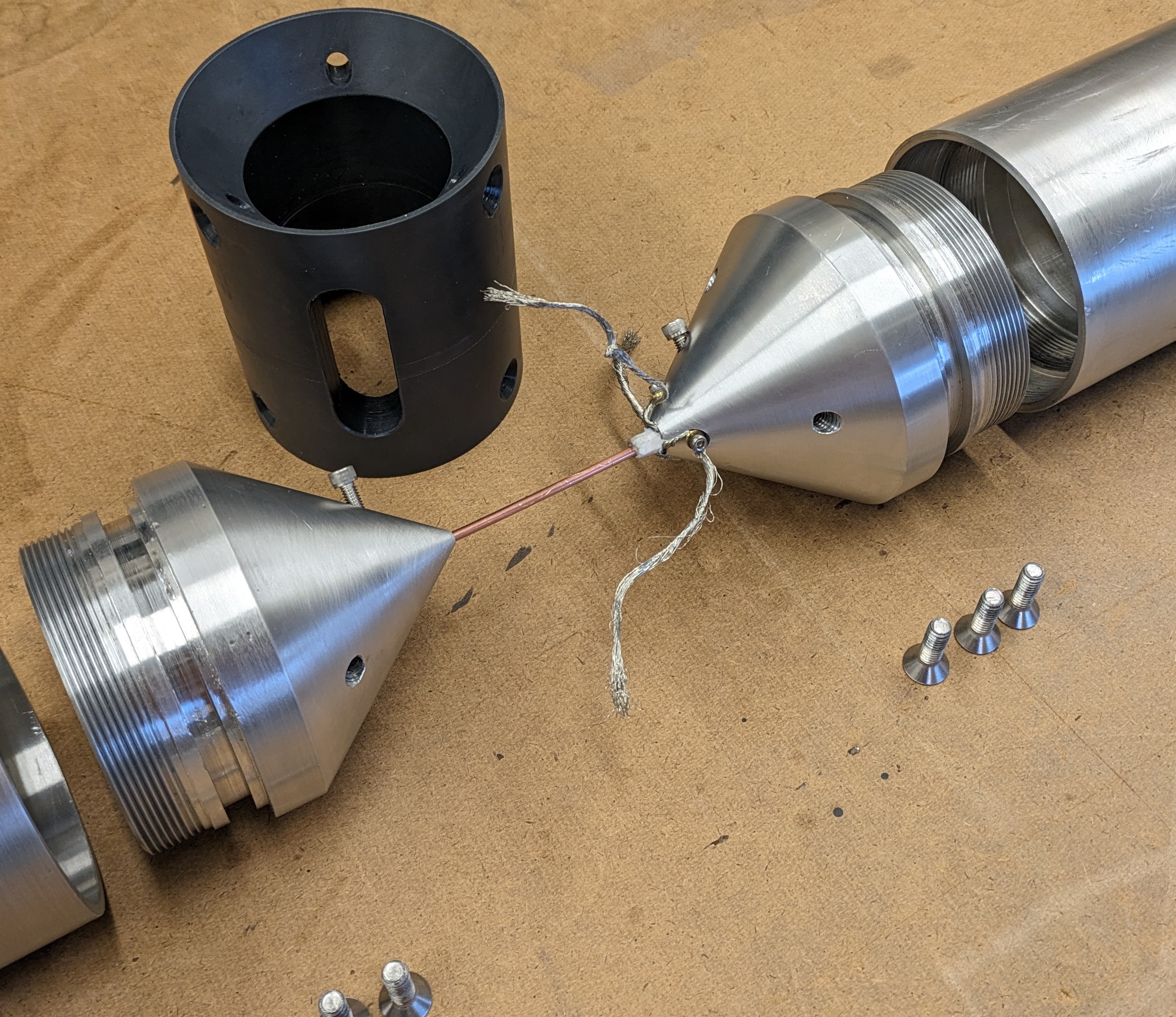}
    \includegraphics[angle=90,width=0.14\linewidth]{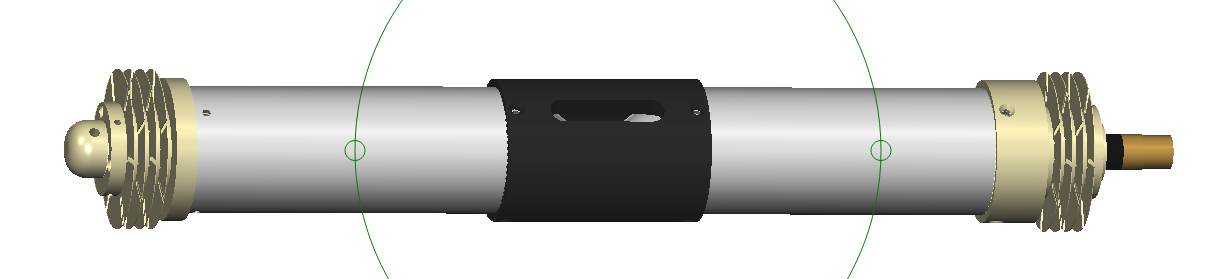}
    \caption{Photograph of central feed components of GUNK PVA, used for refractive index measurements at Summit Station, Greenland (left) and CAD rendition of fat dipole antenna, showing endcaps at either end, and N-male connector for conveying signal (top).}
    \label{fig:SPUNK_photos}
\end{figure}

For pressure sealing, Buna-N 2-335 O-rings were selected (although, in retrospect, more expensive silicone o-rings would have yielded superior low temperature performance), with Nylon 6/6  used for the caps, based on their long heritage with the RICE dipoles. Nylon is also strong (for maintaining pressure) and reasonably machinable.
Drawings and pictures of the SPUNK antennas, including details on dimensions and shape,
are presented in Figs. \ref{fig:SPUNK_drawings} and \ref{fig:SPUNK_photos}.

\subsection{Antenna Radio-Frequency characteristics}
\begin{figure}
    \centering
    \includegraphics[width=0.8\linewidth]{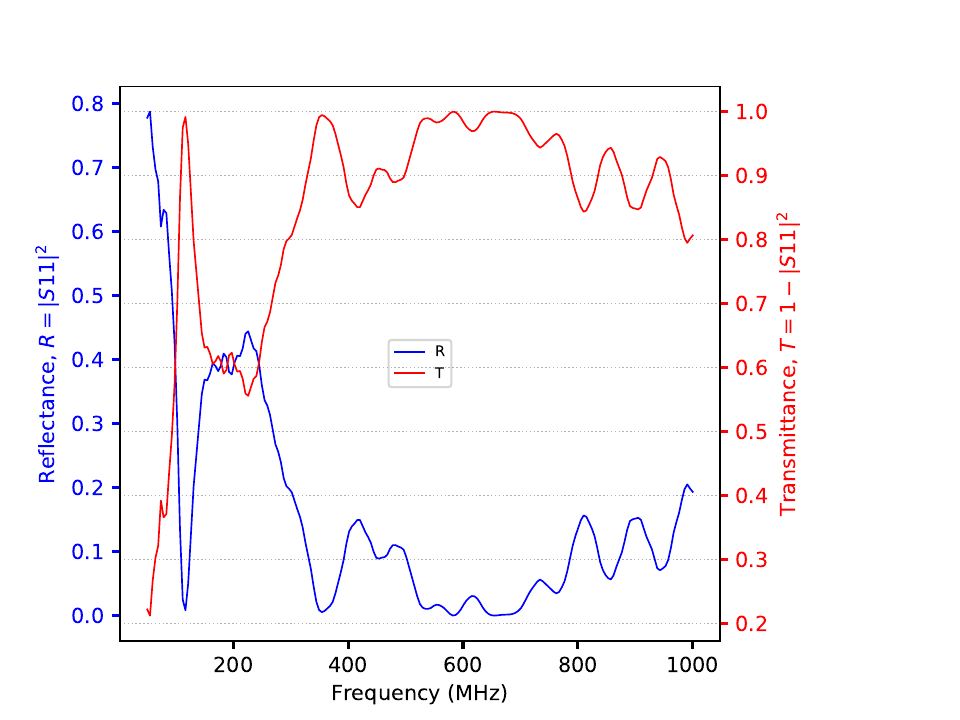}
    \caption{Antenna performance, as tested in University of Kansas (KU) anechoic chamber. `Transmittance' (right scale) displays the fraction of signal power broadcast by the antenna, relative to the power input at the feedpoint.}
    \label{fig:antresponse}
\end{figure}

After construction (but without the signal generator connected to the feedpoint), verification of radio-frequency performance was required.
The SPUNK antenna was tested in the Center for Remote Sensing and Integrated Systems (CReSIS) anechoic chamber, located on the University of Kansas, and rated at frequencies down to 30 MHz; this afforded an opportunity to perform precise characterization of antenna RF capabilities. Radio-frequency performance is typically measured via Vector Network Analyzer `S-parameters' data, which quantify the efficiency of an antenna to radiate input signal into the environment (in SEND mode) or, equivalently, the efficiency of an antenna to convert input radiation energy into currents along a transmission line attached at the antenna feed point (in RECEIVE mode). Although the textbook `ideal' dipole of length L resonates at a single frequency given by f=2$c_0$/L, the `fat' dipole design of the SPUNK antenna enhances the bandwidth and therefore also improves the signal response for UHEN-induced Askaryan radiation. Figure
\ref{fig:antresponse} displays the in-air calibration data obtained for the SPUNK antenna in the University of Kansas anechoic chamber, showing high transmittance (small reflection coefficient at the feedpoint) over a wide frequency range.

\section{Field Performance}

\begin{figure}
    \centering
    \includegraphics[width=0.5\linewidth]{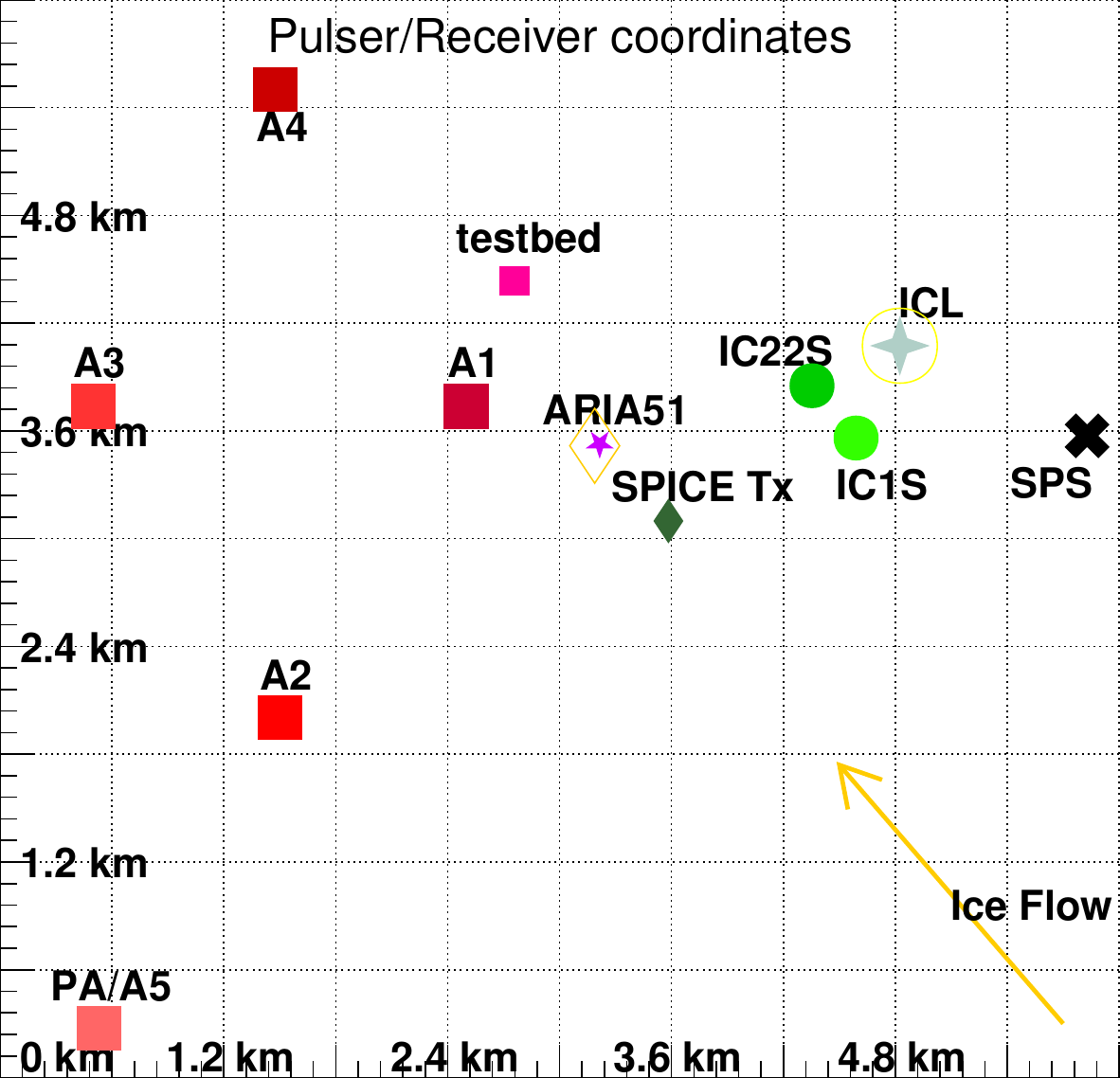}
    \caption{Map showing primary landmarks associated with calibration transmitter pulsing. Indicated in the Figure are the five ARA radio receiver stations active at the time of the SPUNK PVA calibration campaign in December, 2018, ARIANNA station 51 (which used detected SPUNK signals to measure polarization reconstruction precision of ARIANNA radio array), the Icecube Laboratory (ICL, housing electronics and data acquisition hardware for the IceCube experiment at South Pole, as well as location of IceCube iceholes 1 and 22, each of which contain separate radio-frequency calibration transmitters (stationary) deployed in December 2011 and also the original ARA `testbed', active 2011-12 and used to validate ARA detection scheme and electronics. `SPICE Tx' indicates location of 1750-m deep SPICE ice borehole, used for primary measurements referred to in this article. Note direction of ice flow in the Figure, allowing comparison of signal propagation wave speed both parallel to (in the direction of ARA stations A1 and A4, e.g.) and also perpendicular to (in direction of ARA stations A2 and PA/A5, e.g.) ice flow direction. The wave speed asymmetry allows us to measure the birefringent properties of the ice crystal fabric.}
    \label{fig:SPmap}
\end{figure}

\begin{figure}
    \centering
    \includegraphics[width=0.49\linewidth]{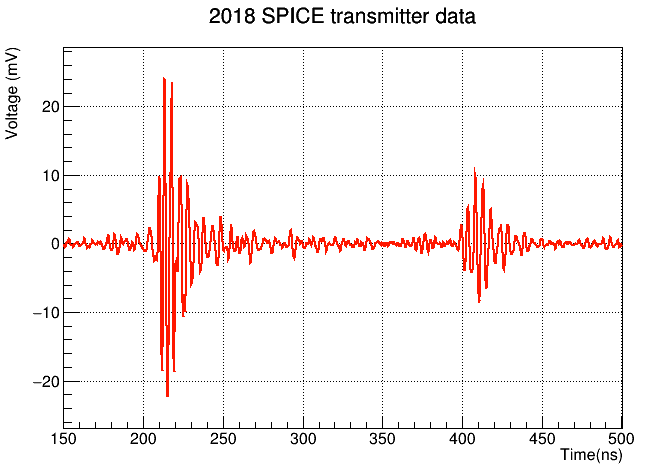}
        \includegraphics[width=0.49\linewidth]{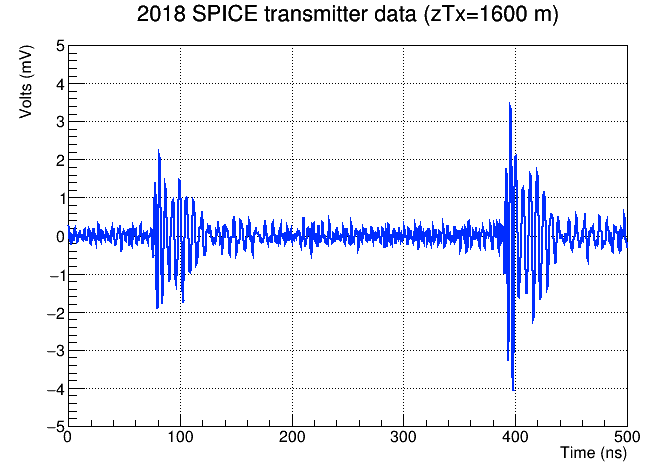}
    \caption{Sample single-channel event waveform captures for transmitter at a depth of 820 m broadcasting to ARA station A2 (left) and also at depth of 1600 m broadcasting to ARA station A4 (right). Note the time difference, in each panel, between the first Direct (`D') pulse and the second `Refracted' (R) pulse. The time delay between D and R signal arrival times, coupled with knowledge of the receiver depth, can be translated into an estimate of the range to the source interaction point and has also been used to determine the refractive index profile with high precision. Note the difference in relative signal strengths for D:R at z=820 m vs. z=1600 m, due to trajectory-dependent differences in ice focusing.}
    \label{fig:signals}
\end{figure}

The 1750-m deep borehole drilled for the South Pole Ice Core Experiment (SPICE) provided an opportunity to field test the pulser+antenna; a map showing local landmarks is presented in Figure \ref{fig:SPmap}. Over the course of 9 days (12/23/18--12/31/18), the PVA antenna was lowered, via winch, into the SPICE borehole eight times; during that time, signals were broadcast to the ARA RF receiver array. Each of the 5 ARA stations triggered on pulses from the PVA transmitter, at distances ranging from 1.5 to 5 km. Data collected during the SPICE campaign were used to measure and model {\it in situ} birefringence \cite{jordan2020modeling, connolly2022impact, flaherty2024polarization,allison2020long}, the radio-frequency attenuation length \cite{allison2020long} and the refractive index profile \cite{KennyPaper}. Fig. \ref{fig:signals} illustrates the high-fidelity signals captured by one of the ARA stations, with the antenna at a depth of 1600 m (right panel) and therefore subjected to pressures approximately 150$\times$ atmospheric. 

\subsection{Physics Results}

As an example of the science afforded by the SPUNK calibration campaign, we detail measurements of the real portion of the dielectric permittivity of cold polar ice at the South Pole. The  refractive index profile (RIP) $n(z)$ is of fundamental importance to radio UHEN experiments. Since a changing index of refraction, by Fermat's Least-Time Principle, results in curved, rather than rectilinear RF ray trajectories, it is essential to understand the firn index of refraction profile \citep{RNOG_Philipp} where radio receivers have been deployed, in order to reconstruct the geometry of neutrino interactions for UHEN experiments. Curved ray paths through the firn were calculated analytically 40 years ago \citep{rasmussen1986refraction}, with particular application to ice thickness calculations from polar radar surveys. 

\begin{figure}    \centering    \includegraphics[width=0.8\linewidth]{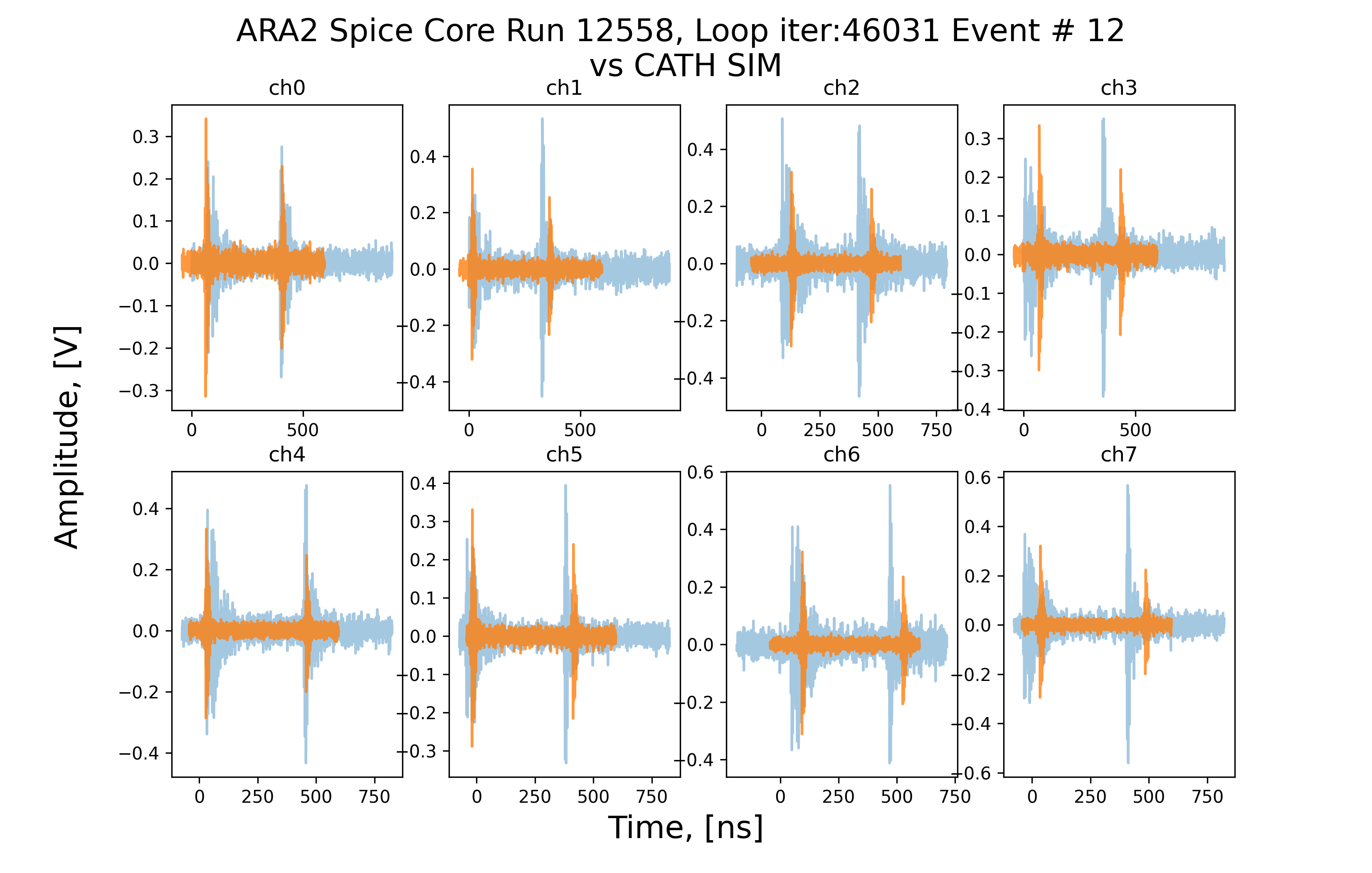}    \caption{Comparison between measured voltage(time) traces (blue) vs. simulated voltage traces (orange) for 8 receiver channels of the Askaryan Radio Array (South Pole) receiving signals from a transmitter deployed into the SPICE core hole. The refractive index profile has been tuned in simulation to give time delays between observed Direct and Refracted signals consistent with data \citep{alisa_nozdrina_2024_thesis}.}    \label{fig:trace-spice}\end{figure}

For a signal generated in a bulk medium, with constant refractive index $n_1$ proximal to a boundary with a second medium having refractive index $n_2$, signals generated in medium 1 and propagating to a receiver also in medium 1 can have two allowed physical trajectories, one of which is direct, and the other reflecting at the boundary interface. If medium 1 has a variable refractive index (such as firn ice), rather than reflecting at the boundary, a refracted signal path will connect transmitter to receiver (Fig. \ref{fig:pictogram}), resulting in two detected signals. The SPUNK PVA was deployed into the South Pole Ice Core Experiment (SPICE) borehole in December, 2018, and broadcast to 2--5 km distant receivers of the ARA experiment. The received signals clearly showed this characteristic double-pulse signature, and also showed agreement with expectations based on our SPICE-derived $n(z)$ model (Fig. 
\ref{fig:trace-spice}). 

The time delay between the Direct and Refracted/Reflected signals (`dt(D,R)') integrates over the ray trajectory and therefore can be used to discriminate between different refractive index models \citep{KennyPaper}.
Conversely, if the RIP is known and the zenith angle of an incident signal measured, one can estimate the distance to a neutrino interaction point simply from dt(D,R) measurements. Note that for a variable RIP, some signal-receiver geometries, particularly for large lateral-separation cases where both the signal production point, and the radio receiver, are both in the firn, are disallowed, resulting in a so-called `shadow zone' \citep{barwick2018observation}. Neutrino interactions within the zone are therefore inaccessible to radio receiver arrays such as ARA, resulting in a reduction in the `effective volume' ($V_{eff}$) over which neutrino detections are possible. The quantitative loss in $V_{eff}$ is assessed using simulations, which find that sub-percent accuracy is desired for an accurate (at the 10\% level) estimate of the loss in sensitive volume due to shadowing. 

Data from the 2018-19 transmitter calibration campaign have, thus far, been seminal in several publications, including:
\begin{enumerate}
\item As outlined above, high precision measurement of the real portion of the ice dielectric permittivity, as a function of depth into the ice, based on measurements of the difference in time between Direct (`D') signal trajectories vs. Refracted/Reflected ('R') signal trajectories (dt(D,R)); R trajectories sample (typically) shallower ice and therefore are more useful in constraining the refractive index over the depth interval where n(z) is changing most rapidly\cite{allison2019measurement,KennyPaper}. The variation in RIP determines the shape and extent of the so-called `shadow zone', which is the main determinant of neutrino sensitivity at the lowest detectable energies.
\item Extraction of the imaginary portion of the ice dielectric permittivity $\varepsilon''$\cite{allison2020long}: The imaginary component of the permittivity determines the attenuation length (`extinction coefficient') of RF signals propagating through cold polar ice. This, in general, can be measured in one of two ways -- an `absolute' measurement via the Friis Equation, requiring that both transmit and receive antennas be absolutely calibrated, or a `relative' measurement, for which identical receiver antennas, at separation distances $\delta$r varying by at least 1 km, allow an extraction of $L_\alpha$ from the relative signal strengths measured for the same transmit signal, and assuming that ice attenuation is given by exp(-$\delta$r/$L_\alpha$).
\item Estimation of the tensorial component of the dielectric permittivity (birefringence)\cite{jordan2020modeling,connolly2022impact}. The planar hexagonal shape of ice crystals and the resulting asymmetry of hydrogen bonds leads to a spatial asymmetry in the electromagnetic wave speed (resulting from an asymmetry in the refractive index $\delta$n) as a function of polarization ${\hat p}$ and wave vector ${\hat k}$, relative to the long axis of the crystal, aka `birefringence'. Spatial asymmetries in the ice sheet result from the vertical gravitational stress and the horizontal strain induced by the longitudinal flow of the ice sheet outwards towards the continental coast (of either Greenland or Antarctica).   
Inputting laboratory measurements of ${\hat p}$ and ${\hat k}$, coupled with {\it in situ} measurements of the crystal orientation fabric (COF), as derived from extracted ice cores, yields predictions for signal arrival time differences for signals broadcast parallel vs. perpendicular to the local ice flow direction.  
\item Measurement of polarization resolution for neutrino-like signals\cite{flaherty2024polarization,anker2020probing}. Compact radio receiver arrays are advantageous in that most (if not all) of the antennas will typically be illuminated within the 2-3 degree transverse thickness of the Cherenkov cone. However, for typical geometries, the cone extends up to one km in extent, such that the cone is not fully imaged, and the incident neutrino momentum vector must be reconstructed from other observables. Since the electric field associated with a Cherenkov cone is polarized transverse to the circular cone front, a precise polarization measurement can be used to unambiguously calculate the direction of the incident neutrino  (see Fig. \ref{fig:pictogram}). The ratio of the signal voltage parallel to the long-axis of the SPUNK antenna relative to signal voltage transverse to the long-axis (i.e., `{\tt VPol:HPol}') can thus be used to infer the neutrino momentum direction.
\end{enumerate}
Such measurements demonstrate the unique science potential offered by an in-ice antenna capable of broadcasting high amplitude signals at km-scale depths.

\section{Conclusion and Summary}
The fact that neutrinos are measured at nearly-horizontal, and/or sub-horizon incidence angles, coupled with the known anisotropy of ice dielectric response, similarly necessitates
transmitter calibration campaigns with corresponding geometries, probing sources at depths up to 2 km within the ice target. However, the narrow cylindrical form factor of typically drilled ice-holes necessitates largely vertical transmitter antennas with high bandwidth (comparable to signals from neutrinos) and active electronics capable of standing off pressures of $\sim$200 bar. We have constructed a custom transmitter antenna to satisfy these requirements, with a demonstrated record of in-field performance. 

Following the calibration work detailed herein, a future campaign, with transmitter immersed in the GISP-2 hole at Summit Station, is being contemplated, provided access to the hole can be secured. That unit would be a copy of the device described here, tailored to the larger width of the GISP-2 hole, and therefore affording greater RF bandwidth; such measurements can hopefully be made in the summer of 2026 or 2027.

\section{Acknowledgments}
 We thank the
Raytheon Polar Services Corporation, Lockheed Martin and the Antarctic Support Contractor for field support
and enabling our work on the harshest continent. We
are thankful to the National Science Foundation (NSF)
Office of Polar Programs and Physics Division for funding support through the IceCube
EPSCoR Initiative (Award ID 2019597) and Award 2013134. We thank the National Science Foundation for their generous support through Grant NSF
OPP-1002483 and Grant NSF OPP-1359535.

\bibliographystyle{unsrt}
\bibliography{cas-refs,mybib}

\end{document}